\documentclass[runningheads]{llncs}

\usepackage{graphicx}
\usepackage{bm}
\usepackage{hyperref}
\usepackage{enumerate}
\usepackage{amsmath,amsfonts}
\usepackage{amssymb}
\usepackage{algorithm}
\usepackage{color}
\usepackage{listliketab}
\usepackage{algpseudocode}
\usepackage{adjustbox}
\usepackage{makecell}
\usepackage{array}
\usepackage{cite}
\newcommand{\matr}[1]{\mathbf{#1}}

\begin{document}

\title{Reconstruction of Demand Shocks \\ in Input-Output Networks}

\author{
Chengyuan Han\inst{1,2}
\orcidID{0000-0001-5220-402X}
\and
Johannes T\"obben\inst{3,4}
\orcidID{0000-0001-7059-3612}
\and
Wilhelm Kuckshinrichs\inst{1}
\orcidID{0000-0002-5181-1298}
\and
Malte Schr\"oder\inst{5}
\orcidID{0000-0001-8756-9918}
\and
Dirk Witthaut\inst{1,2}
\orcidID{0000-0002-3623-5341}
}
\authorrunning{C. Han et al.}

\institute{
Forschungszentrum J\"ulich, Institute for Energy and Climate Research (IEK-STE), 52428 J\"ulich, Germany
\and
Institute for Theoretical Physics, University of Cologne, K\"oln, 50937, Germany
\and
Gesellschaft für Wirtschaftliche Strukturforschung,
49080 Osnabrück, Germany
\and
Potsdam Institute for Climate Impact Research, Social Metabolism \& Impacts, 14412 Potsdam, Germany 
\and
Chair for Network Dynamics, Center for Advancing Electronics Dresden (cfaed) and Institute of Theoretical Physics, Technical University of Dresden, 01062 Dresden, Germany}

\maketitle             

\begin{abstract}
Input-Output analysis describes the dependence of production, demand and trade between sectors and regions and allows to understand the propagation of economic shocks through economic networks. A central challenge in practical applications is the availability of data. Observations may be limited to the impact of the shocks in few sectors, but a complete picture of the origin and impacts would be highly desirable to guide political countermeasures. In this article we demonstrate that a shock in the final demand in few sectors can be fully reconstructed from limited observations of production changes. We adapt three algorithms from sparse signal recovery and evaluate their performance and their robustness to observation uncertainties.  

\end{abstract}

\section{Introduction}
Input-Output (IO) analysis, developed by Wassily W. Leontief, enables the quantitative analysis of the dependence of production, resource requirements and trade between sectors and regions in economic networks \cite{Miller2009}. The central quantity in IO analysis is the Input-Output matrix, describing the underlying inter-dependencies among the sectors or regions (required inputs from one sector or region to another) in terms of a linear mapping \cite{Miller2009}. One central application of Input-Output analysis is to understand and predict the influence of economic shocks. For instance, it predicts how the production in different sectors or regions reacts to a sudden change of the final demand. But what happens if a shock is unknown and we only observe its consequences? 

The availability and quality of economic input data is a central challenge for the assessment of economic shocks and for Input-Output analysis in general. Natural and man-made disasters, for example storms, floods or terrorist attacks, cause shocks to both the demand and supply sides in certain sectors or regions. In many cases, it is only possible to measure the impacts of a disaster on a few sectors directly, but knowing the indirect impacts on other sectors is of great importance for designing policies that aim at enhancing the economy's resilience \cite{hallegatte2008adaptive,okuyama2014disaster,oosterhaven2017wider}. Hence, the application of powerful techniques for data analysis and reconstruction is central in IO analysis.

In this contribution, we analyze to which extend demand shocks can be reconstructed from limited and potentially noisy observations of production changes. This problem cannot be exactly solved in general, unless additional structural information is available. We demonstrate that shocks remain reconstructable if they are sparse, i.e.~if the initial shock is limited to only few sectors or regions. This case is of high relevance, as political decisions, strikes or man-made disasters often affect only few sectors directly, while natural disaster often take place in a limited geographical region. We adapt  different methods from the theory of sparse signal recovery \cite{Tropp&Wright} and evaluate their performance and robustness to observation noise for IO network data from the World Input-Output Database \cite{WIOD}.

\section{Fundamentals and Notations}
\label{sec:fundamentals}

We briefly review the fundamentals of Input-Output analysis analysis following \cite{Miller2009}. In IO analysis, the economic system is divided into $S$ sectors. The total output (production) of the sector $i$ is denoted as $x_i$ and is either sold to other sectors as input for their production or sold to final consumers (households, governments, capital formation and exports), resulting in the balance equation \cite{Miller2009}:
\begin{equation}
    \label{eq:xi=zfi}
    x_i = \sum_{j=1}^{S} Z_{ij} + f_i \,.
\end{equation}
Here, $f_i$ is the final demand for output of sector $i$, and $Z_{ij}$ denotes the inter-industry sales from sector $i$ to sector $j$, where all quantities as measured in equivalent monetary units. To describe the interdependence between sectors $i$ and $j$, Input-Output Analysis assumes fixed production processes and a constant amount $A_{ij}$ of input from sector $j$ required for a unit of output in sector $i$, such that $A_{ij} = {Z_{ij}}/{x_j}$.
We can then rewrite equation \eqref{eq:xi=zfi} as 
\begin{equation}
    \label{eq:xi=afi}
    x_i = \sum_{j=1}^{S}( A_{ij} x_j )+ f_i.
\end{equation}
and recast this equation in vector form
\begin{align}
     \vec{x} =\matr{A} \vec{x} + \vec{f} 
     \qquad  \Leftrightarrow \qquad 
    (\matr{I}-\matr{A})\vec{x}=\vec{f},
    \label{eq:1-A}
\end{align}
where $\matr{I}$ is the $S\times S$ identity matrix. The column sums of $A$ satisfy $\sum_{j=1}^S A_{ij} < 1$ for all $i$, such that one unit of output in sector $i$ requires less than one unit of total input. Under this assumption $\matr{I}-\matr{A}$ is diagonally dominant and thus invertable. This allows us to reverse equation~\eqref{eq:1-A} to express $\vec{x}$ in term of $\vec{f}$ as
\begin{equation}
    \label{eq:Lf}
    \vec{x} = (\matr{I} - \matr{A}) ^ {-1} \vec{f} = \matr{L} \vec{f},
\end{equation}
where $\matr{L}=(\matr{I}-\matr{A})^{-1}$, known as the \textbf{Leontief inverse} or the \textbf{total requirements matrix}. 

\section{Reconstruction of shocks}

\subsection{Sparse shocks and limited observability}

Input-Output Analysis describes how the demand drives the production across economic sectors. In particular it allows to assess the impact of changes of the demand: Assume that an external shock affects the final demand, $\vec f \rightarrow \vec f + \Delta \vec f$. On short timescales, we assume that the matrix $\textbf{A}$ remains unaffected, as it mainly depends on production technology. Still, due to the dependence of the sectors on each other, the exogenous shock will affect the output of all sectors such that $\vec x \rightarrow \vec x + \Delta \vec x$. Due to the linearity of Eq.~(\ref{eq:Lf}) the changes of production and demand are also related by $\Delta \vec x = \vec L \, \Delta \vec f$. Even if we cannot obtain direct information on the shock, we can in principle reconstruct the shock as $\Delta \vec f = \vec L^{-1} \, \Delta \vec x$ from observations of the total output changes. 

However, situations arise, were we may not be able to obtain full information on the output changes $\Delta \vec x$. In this article we consider what happens if our knowledge of $\Delta \vec x$ is limited: let $K_x$ denote the set of sectors where information about the production changes $\Delta x_i$ is available. Instead of the full system, we then only have the following linear system of equations relating the initial shock $\Delta\vec{f}$ to the observations:
\begin{equation}
\label{eq:partx}
    \sum_{j=1}^S L_{ij} \Delta f_j = \Delta x_i,\quad \text{for} \quad i\in K_x.
\end{equation}
If the set $K_x$ does not include all sectors, this equation is underdetermined and thus not uniquely solvable. Hence, an exact reconstruction of the initial shock $\Delta \vec f$ becomes impossible in general. However, we can overcome this problem if we have additional \emph{structural information} about $\Delta\vec{f}$ that we can exploit. If we know that only a few sectors were disturbed initially, then just a few entries of $\Delta\vec{f}$ are non-zero. We can thus assume that the true solution of the underdetermined equation is obtained by minimizing the number of non-zero entries:
\begin{equation}
    \min\|\Delta\vec{f}\|_0\quad \text{such that }\sum_{j=1}^S L_{ij}\ \Delta f_j = \Delta x_i\quad \text{for} \quad i\in K_x,
    \label{eq:l0min}
\end{equation}
where $\|\cdot\|_0$ denotes the number of non-zero elements in a vector. This problem is NP-hard to solve in general such that a direct solution is not achievable. However, the last decade has witnessed significant progress in the development of efficient methods for an indirect solution applicable to many cases of practical importance \cite{marques2018review, timme2014revealing}. A review of different approaches can be found in \cite{Tropp&Wright}.

\subsection{Sparse reconstruction methods}
\label{sec:signalrec}

In this article we test the capability of three algorithms to reconstruct sparse shocks of the demand in an IO network:
\begin{enumerate}
\item 
Convex relaxation ($\ell_1$ minimization): The optimization problem (\ref{eq:l0min}) is hard to solve since $\|\cdot\|_0$ is non-convex. Fortunately, in many cases one can still get the correct solution by replacing the non-convex pseudo-norm $\|\cdot\|_0$ by the convex norm $\|\cdot\|_1$, greatly simplifying the optimization problem \cite{Donoho, Candes}.  
\item
The Orthogonal Matching Pursuit (OMP) approach is based on the following observation \cite{PatiOMP,davisOMP}. If the vector $\Delta \vec f$ is sparse, then only few columns of the matrix $\vec L$ enter the product $L \Delta \vec f$. Hence OMP tries to approximate the signal $\Delta x$ by a superposition of only few columns of $\vec L$, which are chosen one-by-one to reduce the error of the approximation as much as possible in each step.
\item 
The Compressive sampling matching pursuit (CoSaMP) extends and improves OMP in several ways \cite{needellCoSaMP}. For instance, several new columns of $\vec L$ can be chosen in each step. In particular, instead of choosing new columns one-by-one, the algorithm adds several columns in each step and subsequently removes the ones that contribute the least to a correct approximation.
\end{enumerate}

\subsection{First Example}

We illustrate the problem of reconstructing an initial shock using an elementary example from \cite{Miller2009}, Section 2.3.4, describing the US economy on coarse scales with only $S=7$ sectors. The IO matrix $\matr{A} \in \mathbb{R}^{7 \times 7}$ is given by
\begin{equation}
     \matr{A} = 
     \begin{bmatrix}
     0.2008 & 0.0000 & 0.0011 & 0.0338 & 0.0001 & 0.0018 & 0.0009 \\
     0.0010 & 0.0658 & 0.0035 & 0.0219 & 0.0151 & 0.0001 & 0.0026 \\
     0.0034 & 0.0002 & 0.0012 & 0.0021 & 0.0035 & 0.0071 & 0.0214 \\
     0.1247 & 0.0684 & 0.1801 & 0.2319 & 0.0339 & 0.0414 & 0.0726 \\
     0.0855 & 0.0529 & 0.0914 & 0.0952 & 0.0645 & 0.0315 & 0.0528 \\
     0.0897 & 0.1668 & 0.1332 & 0.1255 & 0.1647 & 0.2712 & 0.1873 \\
     0.0093 & 0.0129 & 0.0095 & 0.0197 & 0.0190 & 0.0184 & 0.0228
     \end{bmatrix},
\end{equation}
and the Leontief inverse matrix can be derived from equation \eqref{eq:Lf}. In this example, there is a change in the final demand in sector 1 (agricultural items) and sector 4 (manufactured items) due to foreign demand. In particular,
\begin{equation}
\label{eq:dfexample}
    \Delta \vec f =
    \begin{bmatrix}
    1.2, & 0, & 0, & 6.8, & 0, & 0, & 0
    \end{bmatrix}^\top
\end{equation}
(in million dollars), using the symbol $\top$ to denote the transpose of a matrix or vector. This causes a change of output
\begin{equation}
    \Delta\vec{x}=\matr{L} \Delta\vec{f} =
    \begin{bmatrix}
    1.9114, & 0.2444, & 0.0526, & 9.1249,  & 1.2421, & 2.2709,  & 0.2788
    \end{bmatrix}^\top \, .
\end{equation} 
Now suppose we measure all entries of $\Delta \vec{x}$, then we can simply recover the cause of disturbance $\Delta\vec{f}$ by equation \eqref{eq:1-A}. But what happens if we have incomplete information? Say we only have the information
\begin{equation}
    \Delta x_1=1.9114, \quad \Delta x_3=0.0526, \quad \text{and} \quad \Delta x_6=2.2709 \,.
\end{equation}
Assuming $\Delta\vec{f}$ is sparse we can answer this question as discussed above by appling the algorithms introduced in section \ref{sec:signalrec} to approximately solve the optimization problem
\begin{equation}
    \Delta\vec{f}_R = \text{arg}\min_{\Delta\vec{f}}\|\Delta\vec{f}\|_0, \text{ s.t. } \Delta x_1=1.9114, \, \Delta x_3=0.0526, \, \text{and} \, \Delta x_6=2.2709.
\end{equation}
In this example, we find that both OMP and CoSaMP yield the correct solution for $\Delta\vec{f}$ as shown in equation \eqref{eq:dfexample}. But is reconstruction possible in general and which algorithm is most appropriate? To answer this question, we consider a larger, more realistic representation of an IO network and try different algorithms in the next section.

\section{Result}

\subsection{Impact Recovery for the WIOD Dataset}
\label{sec:impact}

To test whether sparse signal recovery is possible in a real-world setting, we analyze the performance of the reconstruction methods mentioned in section \ref{sec:signalrec} for the World Input-Output Database (WIOD) \cite{WIOD}. 

The WIOD provides multi-regional input-output tables from different years to represent the trade between any two sectors in the world. The latest $2014$ table consists of $28$ EU countries, $15$ other major countries, and the "rest of the world" entries to complete the data set. Each country’s economy is divided into $56$ sectors to portray the different industries. Here we consider only the information of the individual sectors, aggregating the input-output-dependency over all countries such that the IO Matrix $\vec{A} \in \mathbb{R}^{56 \times 56}$ with entries corresponding to average inter-industry sales $\left<A_{ij}\right> \approx 10^{-2}$. 

To test the accuracy of the three algorithms mentioned in section \ref{sec:signalrec}, we evaluate the success rates of the reconstruction, varying both the sparsity of the initial shock $\Delta \vec f$, measured in term of the number $s$ of non-zero entries, and the number of observations $n_o$ of the output changes $\Delta \vec x$. For each combination of values $(s,n_o)$, we synthetically generate a large ensemble of $R = 10^2$  test cases as follows. We uniformly randomly select $s$ sectors and choose the entries of $\Delta \vec f$ in these sectors as random values sampled uniformly from the interval $(0,10]$. The remaining entries are set to zero. We then compute $\Delta \vec x = \matr L \, \Delta \vec f$ and uniformly randomly select $n_o$ sectors to be observed. That is, we randomly choose the set $K_x$ such that $|K_x|=n_o$. For these choices we exclude sector 56 from both the shocks and the observations. This sector summarizes the “activities of extraterritorial organizations and bodies” and only contributes to consumption, not production.

For each of these test cases we attempt to reconstruct the initial demand shock via the three algorithms listed in section \ref{sec:signalrec}. That is, we compute a vector $\Delta f_R$ which satisfies the linear constraints (\ref{eq:partx}) and which shall minimize the sparsity $\| \Delta f_R \|_0$. We compare this reconstructed signal $\Delta \vec{f}_{R}$ with the original shock $\Delta \vec{f}$. If $\|\Delta \vec{f}_R - \Delta \vec{f} \|_2 \leq 10^{-5}$, where $\|\cdot\|_2$ is the Euclidean norm, then the reconstruction is considered successful. This procedure is repeated for each of the $R = 10^2$ realizations for each combination of values $(s,n_o)$ to obtain the average success rate.

\begin{figure}[tb]
    \centering
    \includegraphics[width=.95\linewidth]{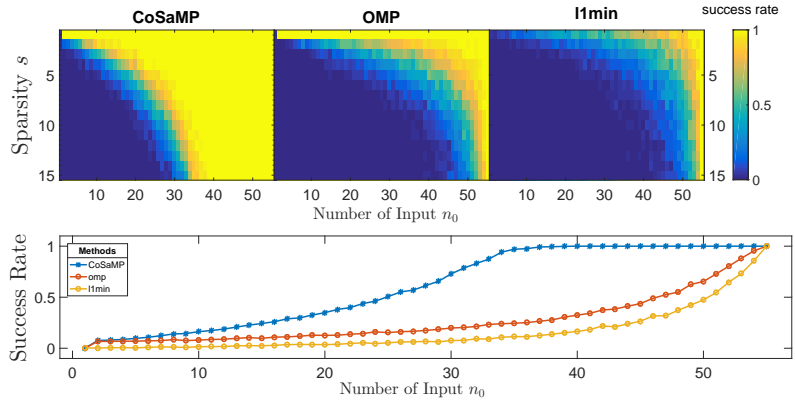}
    \caption{Performance of different algorithms for the reconstruction of demand shocks in IO analysis. Top: the arrays show the success rate of (left to right) CoSaMP, OMP and $\ell_1$ minimization in a color scale plot as a function of the sparsity $s$ of the desired output of the reconstruction $\Delta \vec f$ and the number of observations $n_o$ used as input for the reconstruction.
    Bottom: Success rate as a function of the number of observations $n_o$, averaged over all values of the sparsity $s \in [1,15]$.}
    \label{fig:comparision}
\end{figure}

The numerical results of these tests demonstrate that sparse demand shocks can be reconstructed from limited observations for real-world IO systems (cf.~Figure \ref{fig:comparision}). However, the success rates of the three methods differ vastly. In particular, convex relaxation ($\ell_1$ minimization) performs poorly for the current task. A success rate above 90\% can be achieved only if we have almost complete information about $\Delta \vec x$, i.e.~if $n_0$ is close to the total number of sectors $S=56$. In contrast, CoSaMP shows a very promising performance and outperforms the other algorithms for all values of $s$ and $n_o$. For strongly sparse signals, $s \le 4$, the algorithm has a success rate of 100 \% even if we measure $\Delta x$ for less than half of the sectors. Even for values as high as $s=15$, we find a perfect success rate of 100 \% with a number of inputs $n_o$ well below 40. Because of its superior performance, we focus on the CoSaMP algorithm in the following and analyze its robustness to noisy inputs.

\subsection{Robustness to Measurement Noise}

Trade data can be subject to various forms of inaccuracies. Hence any reconstruction algorithm must be robust to inaccurate or noisy input signals to be useful in practice. To test the robustness of the reconstruction method, we add noise to the observed output changes $\Delta \vec{x}$. Test data is generated as follows. We create an initial shock of sparsity $s$ and select $n_o$ outputs as above. We then select $n_n$ of the $n_o$ observations and add noise to then, drawn independently and uniformly at random from the interval $[0, \sigma]$. The parameter $\sigma \in [0,10^{-1}]$ quantifies the noise strength. The key question is then how this observation noise propagates via the reconstruction algorithm and whether a reconstruction remains possible in principle. To address these questions, we evaluate the Euclidean norm of the difference between the original signal $\Delta \vec{f}$ and the reconstructed signal $\Delta \vec{f}_R$:
\begin{equation}
\label{eq:E_norm}
    E = \| \Delta \vec{f}_R - \Delta \vec{f} \|_2.
\end{equation} 
As above, we define reconstruction be successful if $E < 10^{-5}$ to compute the success rate. We repeat this process for $R = 10^2$ realizations to compute the success rate and the average error. 

\subsubsection{Propagation of uncertainty.}

\begin{figure}[tb]
    \centering
    \includegraphics[width=.95\linewidth,trim={0 0 0 0cm},clip]{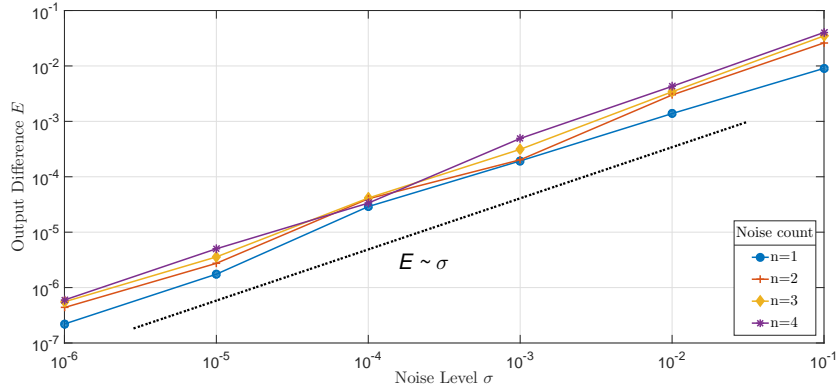}
    \caption{
    Propagation of observation uncertainty ('noise') through the sparse reconstruction. We observe that the error of the reconstructed signal $E=\| \Delta \vec{f}_R - \Delta \vec{f} \|_2$ increases approximately linearly with the noise level. Points correspond to the average over $R = 10^{2}$ repetitions for different values of $n_n$, lines are drawn to guide the eye. The remaining parameters are $s=5$ and $n_o=50$. The dotted line shows a linear scaling $E \sim \sigma$ for comparison.
    }
    \label{fig:outputvsinput}
\end{figure}

Figure \ref{fig:outputvsinput} illustrates the propagation of noise in the reconstruction process. We find that the reconstruction error $E$ increases approximately linearly with the noise level $\sigma$, a finding that is largely independent of the number of noisy observations $n_n$. We conclude that the reconstruction process is robust in the sense that a small amount of noise in the observations causes only a small error in the reconstruction $\Delta f_R$. In particular, a limited amount of noise does \emph{not} render the results of the reconstruction algorithm unfeasible.
 
\subsubsection{Success rate.} 
\begin{figure}[tb]
    \centering
    \includegraphics[width=.95\linewidth]{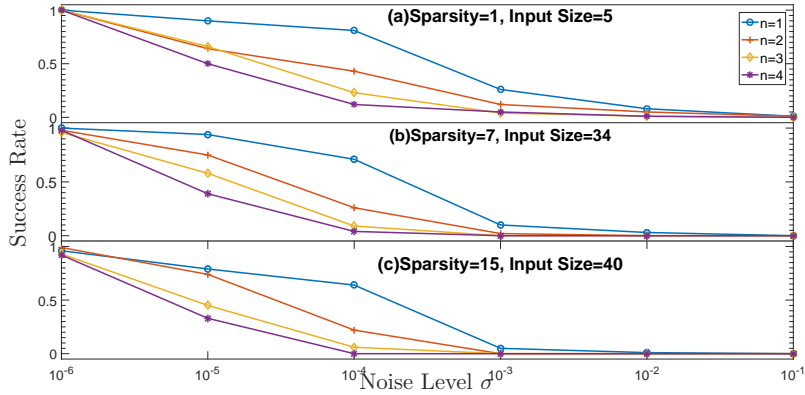}
    \caption{
    Impact of noise on the reconstructability of demand shocks for the WIOD IO network. The panels show the success rate of the reconstruction method as a function of the noise level $\sigma$ for three different parameter settings, $(s,n_o) \in  \{ (1,5), (7,34), (15, 40) \}$ from top to bottom. Point types correspond to different values of $n_n$ and the lines are drawn to guide the eye. A reconstruction is considered successful if 
    $ E = \| \Delta \vec{f}_R - \Delta \vec{f} \|_2 < 10^{-5}$.
    }
    \label{fig:coulev}
\end{figure}

\begin{figure}[tb]
    \centering
    \includegraphics[width=.95\linewidth]{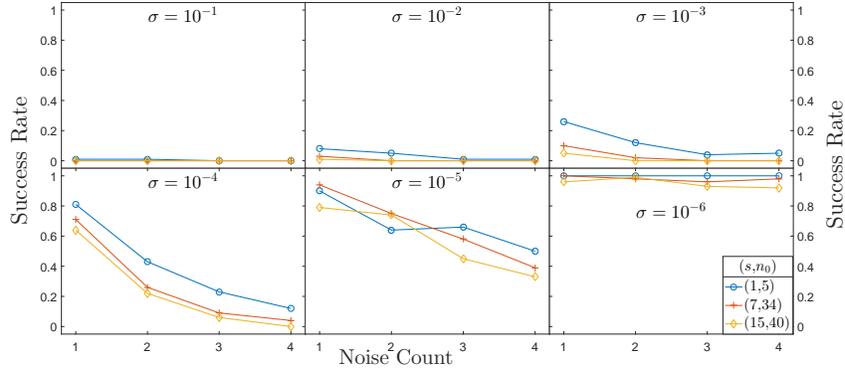}
    \caption{Success rate of reconstruction of the signal $\Delta \vec{f}$ versus the number $n_n$ of measurement subject to noise (the "noise count"). Each plot represents a level of noise $\sigma$. The three lines in each plot represent a different parameter choice of sparsity $s$ and number of observation $n_o$. The success rate of the reconstruction is approximately inversely proportional to the count of the noise $n$ added to the input signal when the noise level $\sigma$ is close to the threshold of reconstructability. Lower sparsity $s$ (blue circles) is more robust than the other two cases for almost all situations.}
    \label{fig:coupts}
\end{figure}

The reconstruction method does not 'amplify' the uncertainty of the observation, but still the performance will be degraded if the noise becomes too strong. To assess these limitations in more detail, we evaluate the success rate as a function of the noise level $\sigma$ and the number of noisy observations $n_n$ in detail in Figure \ref{fig:coulev} and \ref{fig:coupts}. Results are shown for three values $s=1,7,15$, representing cases of low, medium, and high sparsity. In each case, the number of observations $n_o$ was then chosen such that the points are on the boundary of $100\%$ success rate in the noiseless case (compare Fig.~\ref{fig:comparision}). That is, a success rate of 100 \% is found in the noiseless case for the given value of $n_o$, but not for smaller values of $n_o$. This procedure yields the values $(s,n_o) \in \{(1,5), (7,34), (15, 40)\}$.

The presence of observation uncertainty can indeed significantly reduce the success rate as shown in \ref{fig:coulev} - but this depends strongly on the remaining parameters. A limited amount of noise $\sigma$ has only a limited influence on the reconstruction $\Delta \vec f$, but can be enough to increase the error rate $E$ above the threshold value $10^{-5}$defining a successful reconstruction. In particular for the given parameter values $(s,n_o)$ at the boundary of the $100\%$ success region, already weak noise can exceed this threshold. As a consequence, the success rate shown in Figure \ref{fig:coulev} decreases with $\sigma$ and
drops to zero for values between $10^{-4}$ and $10^{-1}$ depending on the remaining parameters. The higher the number of noisy observations $n_n$ the faster the success rate decreases.

The number of uncertain observations $n_n$ can have a strong influence on the success rate if the parameters $(s,n_o,\sigma)$ are such that the $E$ is of the order of the threshold value $10^{-5}$ as shown in Figure \ref{fig:coupts}. For $\sigma=10^{-4}$ the success rate drops by more than $60\%$ when $n_n$ is increased from $1$ to $4$. In other cases, when we are not close to the threshold, the success rate is largely independent of $n_n$

In conclusion, we have demonstrated the robustness of the reconstruction process and we mapped out the consequences of imperfect observation. Uncertainty of the observations ('noise') propagates but does not render the algorithm unfeasible in principle. In practice, it crucially depends on the the tolerable error of the reconstruction whether noise may be problematic or not.

\section{Discussion}

In this contribution we have analyzed the inference of economic shocks from  limited observations in IO networks. We have demonstrated that it is possible to reconstruct a change of the final demand $\Delta f_i$ in few sectors from limited observations of production changes $\Delta x_j$. The key step is to utilize structural information about the initial demand shocks. If shocks emerge from few sectors, the vector of demand changes $\Delta f$ is sparse, enabling the use of advanced methods for sparse signal reconstruction. The best performance was obtained using the Compressive sampling matching pursuit (CoSaMP) algorithm and it was demonstrated that this approach is robust against small uncertainties in the observations. Nevertheless, large uncertainties can be crucial depending on the required accuracy of the reconstructed signals. Even in cases where the reconstruction is not quantitatively successful, this approach may still help to identify which sectors were the source of the initial demand shock.

We note that a successful reconstruction of $\Delta f$ also allows to reconstruct the missing information about the production changes via the relation $\Delta \vec x = \matr L \, \Delta \vec f$. Hence, a reconstruction yields both the origin and the impacts of shocks, which is of great importance to design policies to enhance the resilience of economic networks \cite{hallegatte2008adaptive,okuyama2014disaster,oosterhaven2017wider}. 

\section*{Acknowledgments}

We gratefully acknowledge support from the Helmholtz association (grant no. VH-NG-1025), the German Ministry for Education and Research (BMBF grant no. 03SF0472) and the German Research Foundation (DFG) through the Cluster of Excellence \emph{Center for Advancing Electronics Dresden} (cfaed) and the project 'Bilinear Compressed Sensing'.

\bibliographystyle{splncs04}
\bibliography{ionetwork}

\end{document}